\documentclass[12pt]{article}

\usepackage{graphicx} 
\usepackage{pgfplotstable}
\usepackage{array}
\usepackage{jheppub}
\usepackage{ textcomp }
\usepackage{amsfonts}
\usepackage{amsmath} 
\usepackage{mathrsfs} 
\usepackage{setspace}
\usepackage[utf8]{inputenc} 
\usepackage{comment} 
\usepackage{siunitx}
\usepackage{booktabs}
\makeatletter
\def\@fpheader{\relax}
\makeatother

\DeclareMathAlphabet{\mathbbold}{U}{bbold}{m}{n} 
\newcommand{\be}{\begin{equation}} \newcommand{\ee}{\end{equation}}

\DeclareMathOperator{\Tr}{Tr}

\title{Vector perturbations of Kerr-AdS$\mathbf{_5}$ and the
  Painlevé VI transcendent} 

\author[a,b]{Juli\'{a}n Barrag\'{a}n Amado,}\emailAdd{j.j.barragan.amado@rug.nl}
\author[a]{Bruno Carneiro da Cunha,}\emailAdd{bcunha@df.ufpe.br}
\author[b,c]{and Elisabetta Pallante}\emailAdd{e.pallante@rug.nl}

\affiliation[a]{Departamento de F\'{i}sica, Universidade Federal de Pernambuco,
50670-901, Recife, Pernambuco, Brazil} 
\affiliation[b]{Van Swinderen Institute for Particle Physics and
  Gravity, University of Groningen, 9747 Groningen, Netherlands}
\affiliation[c]{NIKHEF, Science Park 105, 1098 XG Amsterdam, Netherlands} 

\abstract{We analyze the Ansatz of separability for Maxwell equations
  in generically spinning, five-dimensional Kerr-AdS black holes. We
  find that the parameter $\mu$ introduced in \cite{Lunin:2017drx} can
  be interpreted as apparent singularities of the resulting radial and
  angular equations. Using isomonodromy deformations, we describe a
  non-linear symmetry of the system, under which $\mu$ is tied to the
  Painlevé VI transcendent. By translating the boundary conditions
  imposed on the solutions of the equations for quasinormal modes in
  terms of monodromy data, we find a procedure to fix $\mu$ and study
  the behavior of the quasinormal modes in the limit of fast spinning
  small black holes.} 

 
\begin{document}

\maketitle

\section{Introduction}

Black holes in higher dimensions \cite{Emparan:2008eg,Horowitz:2012nnc}
are important to understand aspects of the gauge/gravity
correspondence, with the ultimate goal a better understanding of
theories with non-trivial infrared fixed points. On the other hand, a
better understanding of general relativity is an interesting goal
\textit{per se}, with a clear view on the generic properties and the
features which are special to four dimensions well worth pursuing.

Black hole solutions are particularly distinguished by their
integrable structure. The first example is the
four-dimensional vacuum solution given by the Kerr geometry, which can
be found explicitly even though its isometries -- time translation and
axial rotation -- do not warrant integrability of the equations in
the Liouville sense. The solution was generalized to non-zero
cosmological constant by Carter \cite{Carter:1968ks}, and to higher even
dimensions $D=2n$ by Myers and Perry \cite{Myers:1986un}, 
characterized by $n$ conserved charges. In odd 
dimensions, they were constructed in \cite{Hawking:1998kw} for the
particular case $D=5$ and then, generically, in
\cite{Gibbons:2004js,Gibbons:2004uw}. The 
family of solutions present an integrable set of null congruences, and
the integrability of the solutions themselves can be ascribed to the
existence of higher-rank tensors, satisfying an analogue of the Killing
equation for isometries, the so-called Killing tensors.

This integrable structure, called \textit{hidden symmetries}, allows
not only for the construction of the solutions, but also for
separability of the scalar and spinor wave equations
\cite{Frolov:2017kze}. For spin $1$ fields, however, the situation is
murkier. The separation of Maxwell's equations in four dimensions,
obtained first by Teukolsky \cite{Teukolsky:1972}, is a result of the
existence of a Killing-Yano conserved tensor. In higher dimensions,
the separation was achieved by Lunin \cite{Lunin:2017drx} at the
expense of the introduction of an arbitrary parameter $\mu$. This new
technique was dubbed ``$\mu$-separability'' in
\cite{Frolov:2018eza}. The new parameter is related to the existence of
different polarizations of the electromagnetic and Proca fields
\cite{Krtous:2018bvk,Dolan:2018dqv}, as well as the 
higher $p$-form generalization considered in
\cite{Lunin:2019pwz}. Because of this, the treatment of tensor fields
in these black hole backgrounds is quite different from the scalar
case. The introduction of this extra separability parameter brings in
further questions, related to which  physical requirements should fix
its value, as in the determination of scattering coefficients, angular
eigenvalues and the frequency quasinormal modes.

Coming from a different perspective, the separability of the scalar
wave equation in the subcase of a five dimensional black hole with a
negative cosmological constant -- Kerr-AdS$_5$ -- was tied to the
construction of two flat holomorphic connections in a previous article
by the authors \cite{Amado:2017kao}, related to the solutions of the
angular and radial differential equations. There, the purpose was
solely ``dynamical'': flat holomorphic connections have a residual
gauge-symmetry which allows for solving the connection problem of the
differential equations \cite{Barragan-Amado:2018pxh}.

The residual gauge symmetry, known as ``isomonodromy transformations''
in the theory of ordinary differential equations \cite{Iwasaki:1991}, 
is realized in the angular and radial equations by the presence of an
extra singular point in the Fuchsian equations, whose monodromy
properties are trivial. This extra apparent singularity can be moved
around the complex plane, and the isomonodromy transformation forces a
functional dependence between the position of the apparent singularity
and the positions of the other singularities, which was found to be
the celebrated Painlevé transcendent of the sixth type.

In the scalar case, these extra singularities play an auxiliary role
in the actual solution of the problem: quantities such as scattering
amplitudes and the quasinormal modes depend solely on the monodromy
data. One can then compute them at any point of the isomonodromic flow,
with the coincident point where the apparent singularity merges with
one of the remaining singularities being particularly convenient.

The purpose of the present article is to study the $\mu$-separability
in the particular case of the spin $1$ field in a generic Kerr-AdS$_5$
black hole in order to further elucidate the role of the $\mu$
parameter.  As we will see this is directly related by a Möbius
transformation to the Painlevé transcendent, and parametrizes the
position of the apparent singularity of both the radial and angular
equations. This leads us to the conclusion that the role of $\mu$ in
higher dimensions is different to that in four dimensions, where it
can be eliminated by a change of parametrization in the corresponding
equations.

In the case considered here, the trick of ``deforming'' the Heun
equation by adding an extra, apparent singularity is
mandatory. We will see, however, that the boundary conditions for
angular eigenvalues and quasinormal modes can be written in terms of
monodromy, and hence can be thought of as isomonodromy
invariants. Assuming this invariance, we are able to fix the
parameter $\mu$ through a consistency condition of the isomonodromic
flow in the radial and angular equations. We then proceed to a short
numerical analysis of the solution proposed and close with a short
discussion and prospects.

\section{Maxwell perturbations on Kerr-AdS$\mathbf{_5}$}
\label{Kerr_AdS}

The five dimensional, generically rotating, Kerr-${\rm AdS}_5$ metric
was given in \cite{Hawking:1998kw} 
\begin{multline}
ds^{2} =
-\dfrac{\Delta}{\rho^{2}}\left(dt-
  \dfrac{a_1\sin^{2}\theta}{1-a_1^2}d\phi-
  \dfrac{a_2\cos^{2}\theta}{1-a_2^2}d\psi\right)^{2}+
  \dfrac{\rho^{2}}{\Delta}dr^{2}+\dfrac{\rho^{2}}{1-x^2}
  d\theta^{2} \\
  +\dfrac{(1-x^2)\sin^{2}\theta \cos^{2}\theta}{x^2\rho^{2}}\left[
    \left(a_{2}^{2}-a_{1}^{2}\right)dt+\dfrac{a_1(r^{2}+a_1^{2})}{1-a_1^2}
    d\phi-\dfrac{a_2(r^{2}+a_2^2)}{1-a_{2}^{2}}d\psi \right]^{2}  \\
  +\dfrac{a_1^2 a_2^2}{x^2r^2}
  \left[dt -\dfrac{(r^2 +a_1^2)\sin^{2}\theta}{a_1(1-a_1^2)}d\phi -
    \dfrac{(r^2 +
      a_2^2)\cos^{2}\theta}{a_2(1-a_2^2)}d\psi\right]^{2}, \label{eq:kerrads}    
\end{multline}
where
\begin{equation}
\begin{gathered}
  \Delta =
  \dfrac{1}{r^{2}}(r^{2}+a_1^{2})(r^{2}+a_2^{2})(1+r^2)-2M,
  \quad\quad
  x^2 = a_1^2\cos^2\theta+a_2^2\sin^2\theta \\
  \rho^2= r^2+x^2
\label{eq:eq2}
\end{gathered}
\end{equation}
and $a_1$ and $a_2$ are two independent rotation parameters.
This particular form of the metric allows to define an orthonormal
1-form basis $e^{A}$\footnote{$e^{A} = e^{A}_{\mu}dx^{\mu}$, the
  Lorentz indices run as follows $A=\{0,1,2,3,4\}$.}
\begin{subequations}
\begin{equation}
  e^{0}=\sqrt{\dfrac{\Delta}{r^2+x^2}}
    \left(dt-\dfrac{a_1\sin^{2}\theta}{1-a_1^2}d\phi-
      \dfrac{a_2\cos^{2}\theta}{1-a_2^2}d\psi\right),
  \end{equation}
  \begin{equation}
    e^{1}=\sqrt{\dfrac{r^2+x^2}{\Delta}}dr,
  \end{equation}
  \begin{equation}
    e^{2}=\sqrt{\dfrac{r^2+x^2}{1-x^2}}d\theta,
  \end{equation}
  \begin{equation}
    e^{3}=\sqrt{\dfrac{1-x^2}{r^2+x^2}}\dfrac{\sin\theta
    \cos\theta}{x}\left(\left(a_{2}^{2}-a_{1}^{2}\right)dt+
    \dfrac{a_1(r^{2}+a_1^{2})}{1-a_1^2}d\phi-\dfrac{a_2(r^{2}+a_2^2)}{1-a_{2}^{2}}
    d\psi \right),
  \end{equation}
  \begin{equation}
    e^{4}=\dfrac{a_1 a_2}{r x}\left(dt -\dfrac{(r^2
    +a_1^2)\sin^{2}\theta}{a_1(1-a_1^2)}d\phi - \dfrac{(r^2 +
    a_2^2)\cos^{2}\theta}{a_2(1-a_2^2)}d\psi\right), 
  \end{equation}
\end{subequations}
which, then, allows us to write
\begin{equation}
ds^{2} = -(e^{0})^{2}+(e^{1})^{2}+(e^{2})^{2}+(e^{3})^{2}+(e^{4})^{2}.
\end{equation}
The inverse metric has a similar factorization
\begin{equation}
g^{\mu\nu}\partial_{\mu}\partial_{\nu} =
-(e_{0})^{2}+(e_{1})^{2}+(e_{2})^{2}+(e_{3})^{2}+(e_{4})^{2}, 
\end{equation}
where
\begin{subequations}
\begin{equation}
  e_{0}=\sqrt{\dfrac{1}{\Delta(r^2+x^2)}}
    \dfrac{(r^2+a_1^2)(r^2+a_2^2)}{r^{2}}\left(\partial_{t}+
    \dfrac{a_1(1-a_1^2)}{r^2+a_1^2}\partial_\phi+
    \dfrac{a_2(1-a_2^2)}{r^2+a_2^2}\partial_\psi\right),
  \end{equation}
  \begin{equation}
    e_{1}=\sqrt{\dfrac{\Delta}{r^2+x^2}}\partial_r,
  \end{equation}
  \begin{equation}
    e_{2}=\sqrt{\dfrac{1-x^2}{r^2+x^2}}\partial_\theta,
  \end{equation}
  \begin{equation}
    e_{3}=\dfrac{1}{\sqrt{(1-x^2)(r^2+x^2)}}
    \dfrac{\sin\theta \cos\theta}{x}\left(
    \left(a_{1}^{2}-a_{2}^{2}\right)\partial_t+\dfrac{a_1(1-a_1^{2})}{\sin^{2}\theta}
    \partial_\phi-\dfrac{a_2(1-a_2^2)}{\cos^{2}\theta}\partial_\psi\right),
  \end{equation}
  \begin{equation}
     e_{4}=-\dfrac{a_1 a_2}{r x}\left(\partial_t +\dfrac{(1-a_1^2)}{a_1}
    \partial_\phi+\dfrac{(1-a_2^2)}{a_2}\partial_\psi\right).
  \end{equation}
\end{subequations}

Following \cite{Lunin:2017drx}, to separate the radial and
angular equation for the gauge field, we need to construct a special
frame with a pair of real null vectors, a pair of complex null vectors
and a space-like unit vector orthogonal to all others. They are given
as $\ell,n,m,\bar{m},k$ as follows
\begin{subequations}
  \label{eq:funfbein}
\begin{align}
  \ell
  & =\sqrt{\dfrac{r^2+x^2}{\Delta}}\left(e_{0} +
    e_{1}\right) \nonumber \\
  & =\dfrac{(r^2+a_1^2)(r^2+a_2^2)}{r^{2}\Delta}\left(\partial_{t}+
   \dfrac{a_1(1-a_1^2)}{r^2+a_1^2}\partial_\phi+\dfrac{a_2(1-a_2^2)}{r^2+a_2^2}
    \partial_\psi\right) +  \partial_r,
\end{align}
\begin{align}
  n &=\dfrac{1}{2}\sqrt{\dfrac{\Delta}{r^2+x^2}}\left(e_0
    - e_1\right) \nonumber \\
  & =\dfrac{(r^2+a_1^2)(r^2+a_2^2)}{2r^{2}(r^2+x^2)}\left(\partial_{t}+
    \dfrac{a_1(1-a_1^2)}{r^2+a_1^2}\partial_\phi+
    \dfrac{a_2(1-a_2^2)}{r^2+a_2^2}\partial_\psi\right)-
    \dfrac{\Delta}{2(r^2+x^2)}\partial_r, 
\end{align}
\begin{align}
  m & =\dfrac{1}{\sqrt{2}}\dfrac{r-ix}{\sqrt{r^2+x^2}}
   \left(e_2 + i e_3\right) \nonumber \\
  & = \dfrac{\sqrt{1-x^2}}{\sqrt{2}(r+i
    x)}\left[\partial_\theta +
    i\dfrac{\sin\theta\cos\theta}{x(1-x^2)}\left((a_1^2-a_2^2)\partial_t
    +\dfrac{a_1(1-a_1^2)}{\sin^{2}\theta}\partial_\phi-
    \dfrac{a_2(1-a_2^2)}{\cos^{2}\theta}\partial_\psi\right)\right],
\end{align}
\begin{align}
  \bar{m} & =(m)^*=\dfrac{1}{\sqrt{2}}\dfrac{r+ix}{\sqrt{r^2+x^2}}\left(e_2
                - i e_3\right) \nonumber \\ 
  & = \dfrac{\sqrt{1-x^2}}{\sqrt{2}(r-i x)}\left[\partial_\theta -
    i\dfrac{\sin\theta\cos\theta}{x(1-x^2)}\left((a_1^2-a_2^2)\partial_t
    +\dfrac{a_1(1-a_1^2)}{\sin^{2}\theta}\partial_\phi-
    \dfrac{a_2(1-a_2^2)}{\cos^{2}\theta}\partial_\psi\right)\right],
\end{align}
\begin{align}
  k & =-\dfrac{a_1 a_2}{r x}\left(\partial_t+
      \dfrac{(1-a_1^2)}{a_1}\partial_\phi+\dfrac{(1-a_2^2)}{a_2}\partial_\psi\right).
\end{align}
\end{subequations}
The first four elements of the list $\ell,n,m,\bar{m}$ are null
vectors -- a null tetrad -- whereas $k$ is orthogonal and space-like
unit vector. Now we define the null transformed ``light-cone'' basis
\begin{equation}
  \begin{gathered}
 \ell_+=\ell,  \qquad \ell_-=-\dfrac{2(r^2+x^2)}{\Delta}n, \\
 m_+=\sqrt{2}(r+i x)m, \qquad m_-=\sqrt{2}(r-i x)\bar{m}=(m_+)^*,
 \label{eq:lightcone}
 \end{gathered}
\end{equation}
leaving $k$ unchanged. Now $(\ell_+,\ell_{-})$ do not depend on the
polar angle $\theta$, and $(m_{+},m_{-})$ do not depend on the
radial coordinate $r$.

\subsection{Separation of variables for Maxwell equations}

In a particular coordinate basis $\{x^\mu\}$, the  source-free Maxwell
equations for a massless vector field can be written as
\begin{equation}
  \label{eq:Maxwell}
  \frac{1}{\sqrt{-g}}\partial_{\mu}\left(\sqrt{-g}F^{\mu\nu}\right)=0,
  \quad \textrm{with}  \quad F^{\mu\nu} = \partial^{\mu}A^{\nu} -
  \partial^{\nu}A^{\mu},
\end{equation}
which unfortunately are \textit{not} separable in the background
\eqref{eq:kerrads}. Again following \cite{Lunin:2017drx}, we can achieve
separability by introducing a parameter $\mu$, and define two
classes of solutions, corresponding to two different polarizations,
called electric and magnetic modes: 
\begin{align}
  \label{eq:ansatz}
  &\ell_\pm^{a}A^{(el)}_{a}=\pm\dfrac{\mu r}{\mu \mp i
    r}\ell_{\pm}^{a}\nabla_{a}\Psi,
   \qquad
   m_{\pm}^{a}A^{(el)}_{a}=\pm\dfrac{i\mu
    x}{\mu \pm
    x}m^{a}_{\pm}\nabla_{a}\Psi,
   \qquad k^{a}A^{(el)}_{a}=0;\\ 
  &\ell_\pm^{a}A^{(mgn)}_{a}=\pm\dfrac{1}{r \pm
    i\mu}\ell_{\pm}^{a}\nabla_{a}\Psi, \qquad
    m_{\pm}^{a}A^{(mgn)}_{a}=\mp\dfrac{i}{x \pm
    \mu}m_{\pm}^{a}\nabla_{a}\Psi, \qquad
    k^{a}A^{(mgn)}_{a}=\lambda\Psi, \nonumber \\
  \label{eq:ansatzmag}
\end{align}
where $\Psi$ is a scalar function which, as we will see below, satisfies
a separable equation. We note that the covariant derivative is always
applied to scalars, so they are independent of the Christoffel
connection. Writing 
\begin{equation}\label{eq:solution}
\Psi = e^{-i\omega t+i \tilde{m}_1\phi + i \tilde{m}_2\psi} \Phi(r)S(x),
\end{equation}
we can express the components of the potential $A_a$ explicitly in the
``light-cone'' basis \eqref{eq:lightcone} and \eqref{eq:ansatz}. For
instance, for the electric solution   $A^{(el)}_a$ ($A^{(el)}_\mu =
A^{(el)}_a(\partial_\mu)^a$): 
\begin{subequations}
  \label{eq:electricA}
  \begin{multline}
    A^{\mathrm{(el)}}_{t}=
    \frac{\Psi}{(r^2+\mu^2)}\biggl\{\frac{\mu^2r\Delta}{(r^2+x^2)}
    \frac{\Phi'(r)}{\Phi(r)}+\frac{S'(x)}{S(x)}\frac{\mu^2(r^2+\mu^2)(1-x^2)
      \sqrt{(a_1^2-x^2)(x^2-a_2^2)}}{(r^2+x^2)(x^2-\mu^2)}\\
    +\frac{\mu}{(x^2-\mu^2)} \bigg[\omega(r^2+\mu^2)(x^2-a_2^2)+
    \omega(r^2+a_1^2)(a_2^2-\mu^2)\\
    -a_1\tilde{m}_1(a_2^2-\mu^2)-a_2\tilde{m}_2(a_1^2-\mu^2)\bigg]\biggr\},
  \end{multline}
  \begin{multline}
    A^{\mathrm{(el)}}_r= \frac{i\Psi}{(r^2+\mu^2)}\biggl[r^2
    \mu\frac{\Phi'(r)}{\Phi(r)} -\frac{\mu^2 (r^2 + a_1^2)(r^2 +
      a_2^2)}{r\Delta}\left(\omega-\dfrac{a_1\tilde{m}_1}{r^2+a_1^2}-
      \dfrac{a_2\tilde{m}_2}{r^2+a_2^2}\right)\biggr],
  \end{multline}
  \begin{multline}
    A^{\mathrm{(el)}}_\theta=\frac{i\Psi}{(x^2-\mu^2)}\biggl[\mu x^2
    \frac{S'(x)}{S(x)}
    -\dfrac{\mu^2\sqrt{(a_1^2-x^2)(x^2-a_2^2)}}{(1-x^2)}
    \left(\omega-\frac{a_1\tilde{m}_1}{(a_1^2-x^2)}-
      \frac{a_2\tilde{m}_2}{(a_2^2-x^2)}\right)\biggr], 
  \end{multline}
  \begin{multline}
    A^{\mathrm{(el)}}_{\phi}=-\frac{a_1\Psi}{(1-a_1^2)(r^2+\mu^2)}
    \biggl[\frac{\Delta\mu^{2}r(x^2-a_1^2)}{(r^2+x^2)(a_2^2-a_1^2)}
    \frac{\Phi'(r)}{\Phi(r)}-
    \frac{\mu^{2}(r^2+a_1^2)(r^2+\mu^2)}{(r^2+x^2)}\\
    \frac{(1-x^2)\sqrt{(a_1^2-x^2)(x^2-a_2^2)}}{(a_2^2-a_1^2)(x^2-\mu^2)}
    \frac{S'(x)}{S(x)}+\frac{\mu}{(x^2-\mu^2)}\biggl(a_1\tilde{m}_1(\mu^2-a_2^2)
    \\
    +a_1\tilde{m}_1(r^2+a_2^2)\frac{(a_2^2-x^2)}{(a_2^2-a_1^2)}
    -a_2\tilde{m}_2(r^2+a_1^2)\frac{(x^2-a_1^2)}{(a_2^2-a_1^2)} \\
   +\omega(a_2^2-\mu^2)(r^2+a_1^2)\frac{(x^2-a_1^2)}{(a_2^2-a_1^2)}\biggr)\biggr],
  \end{multline}
  \begin{multline}
    A^{\mathrm{(el)}}_{\psi}= -\frac{a_2\Psi}{(1-a_2^2)(r^2+\mu^2)}
    \biggl[\frac{\Delta\mu^{2}r(a_2^2-x^2)}{(r^2+x^2)(a_2^2-a_1^2)}
    \frac{\Phi'(r)}{\Phi(r)}-\frac{\mu^{2}(r^2+a_2^2)(r^2+\mu^2)}{(r^2+x^2)}
    \times\\
    \frac{(1-x^2)\sqrt{(a_1^2-x^2)(x^2-a_2^2)}}{(x^2-\mu^2)(a_1^2-a_2^2)}
    \frac{S'(x)}{S(x)}+\frac{\mu}{(x^2-\mu^2)}\biggl(a_2\tilde{m}_2(\mu^2-a_1^2)
    \\
    +a_2\tilde{m}_2(r^2+a_1^2)\frac{(x^2-a_1^2)}{(a_2^2-a_1^2)}
    -a_1\tilde{m}_1(r^2+a_2^2)\frac{(a_2^2-x^2)}{(a_2^2-a_1^2)} \\
    -\omega(r^2+a_2^2)(\mu^2-a_1^2)\frac{(a_2^2-x^2)}{(a_2^2-a_1^2)}
    \biggr)\biggr]. 
  \end{multline}
\end{subequations}

Now, the equations for $\Phi(r)$ and $S(x)$ can be written as
two separate equations, coupled by a separation constant $C_m$ and the
parameter $\mu$
\begin{subequations}
\begin{equation}
  \frac{D_{r}}{r}\frac{d}{dr}\left[\frac{rQ_{r}^{2}(\Delta-R)}{D_{r}}
    \frac{d\Phi}{dr}\right]+\left\{\frac{2\tilde{\Lambda}}{D_r}+\frac{R^{2} 
      \tilde{W}_{r}^{2}}{r^{4}Q_{r}^{2}(\Delta-R)} -
    \frac{a_1^2a_2^2D_r}{r^{2}}\tilde{\Omega}^{2} +
    \mu^2C_mD_{r}\right\}\Phi(r)=0,  \label{eq:radial}
\end{equation}
\begin{equation}
  \frac{D}{x}\frac{d}{dx}\left[\frac{Q^{2}H}{xD}
    \frac{dS}{dx}\right]+\left\{\frac{2\tilde{\Lambda}}{D}-\frac{H
      \tilde{W}^{2}}{Q^{2}x^2} +
    \frac{a_1^2a_2^2D}{x^{2}}\tilde{\Omega}^{2} +
    \mu^2C_mD\right\}S(x)=0 \label{eq:angular}, 
\end{equation}
\end{subequations}
and the functions and constants given by
\begin{equation}
  \begin{gathered}
    R = (r^{2} + a_{1}^{2})(r^2 + a_{2}^{2}), \qquad 
     Q_{r}^{2} = \frac{R(1+r^{2}) - 2Mr^{2}}{r^2(\Delta-R)}, \\ D_{r} = 1
     + \frac{r^{2}}{\mu^{2}},  \qquad  \tilde{W}_{r} =  \omega - \frac{\tilde{m}_1
       a_1}{r^{2} + a_1^{2}}-\frac{\tilde{m}_2 a_2}{r^{2} + a_2^{2}},
  \end{gathered}
\end{equation}
\begin{equation}
  \begin{gathered}
    H=(a_{1}^{2} - x^{2})(a_{2}^{2} - x^{2}), \qquad
    Q^{2} = 1-x^{2},  \\
    D = 1 - \frac{x^{2}}{\mu^2}, \qquad \tilde{W} = \omega
    -\frac{\tilde{m}_1
      a_1}{a_1^{2}-x^{2}}-\frac{\tilde{m}_{2}a_2}{a_2^{2} - x^{2}}.  
  \end{gathered}
\end{equation}
\begin{equation}
  \begin{gathered}
    \tilde{\Lambda} =
    \frac{(a^{2}_{1}-\mu^{2})(a^{2}_{2}-\mu^{2})}{\mu^3}\left(\omega
      -\frac{\tilde{m}_1 a_1}{a_1^{2} - \mu^{2}} - \frac{\tilde{m}_2
        a_2}{a_2^{2} - \mu^{2}}\right), \qquad \tilde{\Omega} =
    \omega 
    -\frac{\tilde{m}_1}{a_1} -\frac{\tilde{m}_2}{a_2}. 
  \end{gathered}
\end{equation}
The subscript $m$ in the separation constant $C_m$ is an integer index
and will be discussed in detail in Sec. \ref{sec:formal}.

The equations above determine the electric polarization, in the sense
described in \eqref{eq:ansatz}, for the potential. The corresponding
equations for the magnetic polarizations were 
also worked out in \cite{Lunin:2017drx}, and it is eventually found
that the function $\Psi$ defined through \eqref{eq:ansatzmag} also
satisfies \eqref{eq:radial} and \eqref{eq:angular}, provided the
separation parameter $\mu$ is substituted by $1/\mu$.  Given $\mu$,
the value for $\lambda$ in \eqref{eq:ansatzmag} is fixed to 
$\tilde{\Omega}/\mu$. The details can be checked in
\cite{Lunin:2017drx} -- although we note the slight change of notation
$\omega(\text{there})=-\omega(\text{here})$, $a_{1,2}(\text{there})= 
-a_{1,2}(\text{here})$, $M(\text{there})=2M(\text{here})$ and
$P_0(\text{there})=\mu^2C_m(\text{here})$. We also note
that, because of the periodicity of the coordinates $\phi$ and
$\psi$, we have $\tilde{m}_i=(1-a_i^2)m_i$, with $m_i$ integers.

Explicitly, the radial and the angular equations are
\begin{multline}
  \frac{r^2+\mu^2}{r}\frac{d}{dr}
  \left[\frac{(r^2-r_0^2)(r^2-r_-^2)(r^2-r_+^2)}{r(r^2+\mu^2)}
    \frac{d\Phi}{dr}\right]+
  \biggl\{-\frac{(r^2+\mu^2)}{\mu^2r^2}(
    a_1a_2\omega-(1-a_1^2)m_1a_2\\
    -(1-a_2^2)m_2a_1)^2
   +\frac{(r^2+a_1^2)^2(r^2+a_2^2)^2}{r^2(r^2-r_0^2)
    (r^2-r_-^2)(r^2-r_+^2)}\left(\omega-\frac{a_1(1-a_1^2)m_1}{r^2+a_1^2}- 
    \frac{a_2(1-a_2^2)m_2}{r^2+a_2^2}\right)^2 \\
  -\frac{2(a_1^2-\mu^2)(a_2^2-\mu^2)}{\mu(r^2+\mu^2)}
  \left(\omega-\frac{a_1(1-a_1^2)m_1}{a_1^2-\mu^2}-
    \frac{a_2(1-a_2^2)m_2}{a_2^2-\mu^2}\right)
  + C_m(r^2+\mu^2)\biggr\}\Phi(r)=0.
  \label{eq:radialheun}
\end{multline} 
and
\begin{multline}
  \frac{(\mu^2-x^2)}{x}\frac{d}{dx}
  \left[\frac{(1-x^2)(a_1^2-x^2)(a_2^2-x^2)}{x(\mu^2-x^2)}
    \frac{dS}{dx}\right]+ \biggl\{
  \frac{(\mu^2-x^2)}{\mu^2x^2}(a_1a_2\omega-a_2(1-a_1^2)m_1\\
    -a_1(1-a_2^2)m_2)^2
  -\frac{(a_1^2-x^2)(a_2^2-x^2)}{x^2(1-x^2)}\left(\omega-
    \frac{a_1(1-a_1^2)m_1}{a_1^2-x^2}-\frac{a_2(1-a_2^2)m_2}{a_2^2-x^2}
  \right)^2\\
  -\frac{2(a_1^2-\mu^2)(a_2^2-\mu^2)}{\mu(\mu^2-x^2)}
  \left(\omega-\frac{a_1(1-a_1^2)m_1}{a_1^2-\mu^2}-
    \frac{a_2(1-a_2^2)m_2}{a_2^2-\mu^2}\right)
  +C_m(\mu^2-x^2)\biggr\}S(x)=0
\label{eq:angularheun}
\end{multline}
where now the values $r_+^2,r_-^2$ and $r_0^2$ are defined,
following \cite{Barragan-Amado:2018pxh}, as the roots of $\Delta$, 
\begin{equation}
  \Delta = \frac{(1-r^2)(r^2+a_1^2)(r^2+a_2^2)}{r^2}-2M
  = \frac{(r^2-r_0^2)(r^2-r_+^2)(r^2-r_-^2)}{r^2}.
\end{equation}

\subsection{The radial and angular systems}
\label{sec:maxwelltoheun}

The radial equation \eqref{eq:radialheun} can be brought to a standard
form by making a M\"obius transformation  
\begin{equation}
z = \frac{r^{2} - r_{-}^{2}}{r^{2} - r_{0}^2}, \quad  \textrm{with}
\quad z_0 = \frac{r_+^2 - r_{-}^{2}}{r_{+}^{2} - r_{0}^{2}}, 
\end{equation}
followed by introducing a new radial function regular at horizon and
the boundary, 
\begin{equation}
  \Phi(z) = z^{-\alpha_{-}}(z-1)^{\alpha_{\infty}}(z- z_{0})^{-\alpha_{+}} R(z).
\end{equation}
The exponents $\alpha_{k}$ are related to the monodromy parameters
$\theta_k$ as 
\begin{equation}
\alpha_{k} = \pm \frac{1}{2}\theta_{k}, \quad k=+,-,0 \quad
\textrm{and} \quad \alpha_{\infty} = \frac{1}{2}\left(1 \pm \sqrt{1 -
    C_m}\right), 
\end{equation}
which in turn are given in terms of the physical parameters by
\begin{equation}
\theta_{k} = \dfrac{i}{2\pi}\left(\frac{\omega -
    m_1\Omega_{k,1}-m_2\Omega_{k,2}}{T_{k}}\right), \qquad
\theta_1=-\sqrt{1 - C_m}.
  \label{eq;singlemonorad}
\end{equation}
We note that, just like the scalar case \cite{Amado:2017kao}, and in
the four-dimensional Teukolsky master equation
\cite{CarneirodaCunha:2019tia}, the monodromy parameters $\theta_+$
and $\theta_-$, respectively associated to the outer and inner
horizonts at $r=r_+$ and $r=r_-$ are proportional to the 
variation of the black hole entropy as a quantum of energy $\omega$ and
angular momenta $m_1$ and $m_2$ passes through the horizon.

With these definitions, the radial equation becomes
\begin{multline}\label{eq:deformedradial}
  \frac{d^{2}R}{dz^{2}}
  +\left[\frac{1-\theta_{-}}{z}+\frac{1-\theta_1}{z-1}+
    \frac{1-\theta_{+}}{z-z_0} 
    - \frac{1}{z-z_\star}\right]\frac{dR}{dz} \\
  +\left(\frac{\kappa_{+}\kappa_{-}}{z(z-1)} +
    \frac{z_{0}(z_{0}-1)K_0}{z(z-1)(z-z_0)}
   + \frac{z_\star(z_{\star}-1)K_\star}{z(z-1)(z-z_\star)}\right)R(z) = 0,
\end{multline}
with the parameters as
\begin{subequations}
  \label{eq:accessory_parameters}
  \begin{equation}
    \label{eq:zstar}
    z_\star = \frac{r_-^2+\mu^2}{r_0^2+\mu^2}
\end{equation}
\begin{equation}
  \kappa_{+}\kappa_{-}=\frac{1}{4}((\theta_{-}+\theta_{+}+\theta_1-1)^2-\theta_0^2),
\end{equation}
\begin{multline}
  4z_{0}K_{0}=
  (\theta_{-}+\theta_{+}+\theta_1-1)^2-\theta_0^2-
  2\theta_{-}\theta_1+2\theta_1-2-
  \frac{2(1-\theta_1)\theta_{+}}{(z_0-1)} +
  \frac{\omega^2}{(r_{-}^2-r_{0}^2)} -\\
  \frac{a_1^2a_2^2\tilde{\Omega}^2}{\mu^{2}(r_{-}^2-r_{0}^2)} 
  +\frac{2z_\star\theta_{+}}{(z_0 - z_\star)}+
  \frac{2(z_\star-1)\mu^3\omega}{(r_+^2-r_0^2)(r_-^2-r_0^2)(z_0-z_\star)}+
  \frac{2z_0(1-\theta_1)}{(z_0-1)}+\\
  \frac{2(z_\star-1)a_1^2a_2^2\tilde{\Omega}}{
    \mu(r_+^2-r_0^2)(r_-^2-r_0^2)(z_0-z_\star)}
  -\frac{2(z_\star-1)\mu((a_1^2+a_2^2)\omega-
    a_1(1-a_1^2)m_1-a_2(1-a_2^2)m_2)}{(r_+^2-r_0^2)(r_-^2-r_0^2)(z_0-z_\star)}
  \\
  +C_m+\frac{(z_0-z_\star)}{(z_0-1)(z_\star-1)}C_m,
\end{multline}
\begin{multline}
  4z_{\star}K_{\star}=-\frac{2(z_\star-1)a_1^2a_2^2
    \tilde{\Omega}}{\mu(r_+^2-r_0^2)(r_-^2-r_0^2)(z_0-z_\star)}
  -\frac{2(z_\star-1)\mu^3\omega}{(r_+^2-r_0^2)(r_-^2-r_0^2)(z_0-z_\star)}
  +2\theta_{-}\\
  +\frac{2(z_\star-1)\mu((a_1^2+a_2^2)\omega-
    a_1(1-a_1^2)m_1-a_2(1-a_2^2)m_2)}{
    (r_+^2-r_0^2)(r_-^2-r_0^2)(z_0-z_\star)}-
  \frac{2z_{\star}\theta_+}{(z_0-z_\star)}-\frac{2z_{\star}(1-\theta_1)}{(z_\star-1)}.
\end{multline}
\end{subequations}

The differential equation \eqref{eq:deformedradial} is Fuchsian, with $5$
regular singular points at $z=0,z_0,z_\star,1,\infty$. It is sometimes
called the deformed Heun equation, because, as we will see below, the
singular point at $z_\star$ is \textit{apparent}: the indicial
coefficients are $\{0,2\}$ and, due to an algebraic relation between
the parameters, there are no logarithmic tails\footnote{The name of K. Heun is usually connected to the Fuchsian equation with $4$ regular singular points. The generic differential equation with $5$ regular singular points has no widespread name, although it was associated to F. Klein and M. Bôcher in the classic treatise of E. L. Ince \cite{Ince:1956}.}. Then the monodromy
property of the solution around this point is trivial. The position of this
apparent singularity is related to the parameter $\mu$ by a Möbius
transformation as it can be seen in \eqref{eq:zstar}. Finally,
we note that the deformed Heun equation \eqref{eq:deformedradial}
depends on $\mu$ only through $z_\star,K_0,K_\star$.

The angular equation \eqref{eq:angularheun} can be brought to the same
form \eqref{eq:deformedradial} by the Möbius transformation
$u = (x^2-a_1^2)/(x^2-1)$. The resulting equation is again Fuchsian
with $5$ regular singular points, located at
\begin{equation}\label{eq:ustar}
  u=0, \qquad u=1, \qquad u=u_0=\frac{a_2^2-a_1^2}{a_2^2-1}, \qquad
  u=u_\star=\frac{\mu^2-a_1^2}{\mu^2-1}, \qquad u=\infty ,
\end{equation} 
and the characteristic exponents are
\begin{gather}
\beta_0^\pm=\pm\frac{m_1}{2}, \qquad \beta_1^{\pm}=\frac{1}{2}\left(1
  \pm \sqrt{1 - C_m}\right), \qquad
\beta_{u_0}^\pm=\pm\frac{m_2}{2}, \\ 
\beta_\star =\{0,2\}, \qquad
\beta_{\infty}^\pm=\frac{1}{2}\varsigma = \frac{1}{2}
\left(\omega+a_1m_1+a_2m_2\right).
\label{eq:singlemonoang}
\end{gather}
We can check that, once more, the point at $u=u_\star$ is an apparent 
singularity due to an algebraic relation between the parameters. 

Finally, the angular equation \eqref{eq:angularheun} can be brought to
a canonical form by the transformation
\begin{equation}
  S(u)=u^{m_1/2}(u-1)^{(1-\theta_1)/2}(u-u_0)^{m_2/2}Y(u),
\end{equation}
which leads to the deformed Heun form \eqref{eq:deformedheun},
\begin{multline}\label{eq:deformedangular}
\frac{d^{2}Y}{du^{2}} +
\left[\frac{1+m_1}{u}+\frac{1-\theta_1}{u-1}+\frac{1+m_{2}}{u-u_0} -
  \frac{1}{u-u_\star}\right]\frac{dY}{du} \\  
+\left(\frac{q_{+}q_{-}}{u(u-1)} +
  \frac{u_{0}(u_{0}-1)Q_0}{u(u-1)(u-u_0)}+
  \frac{u_\star(u_{\star}-1)Q_\star}{u(u-1)(u-u_\star)}\right)Y(u) =
0, 
\end{multline}
with the accessory parameters given by
\begin{subequations}
  \begin{equation}
     q_+q_-=\frac{1}{4}((m_1+m_2+1-\theta_1)^2-
     (\omega+a_1m_1+a_2m_2)^2),
   \end{equation}
   \begin{multline}
     4u_{0}Q_0=
     (m_1+m_2+1-\theta_1)^2-(\omega+a_1m_1+a_2m_2)^2
     +2m_1\theta_1-2(1-\theta_1)+
     \frac{2m_2(1-\theta_1)}{(u_0-1)}\\
     +\frac{2u_0(1-\theta_1)}{(u_0-1)}-
     \frac{a_1^2a_2^2\tilde{\Omega}^2}{\mu^2(1-a_1^2)}+
     \frac{\omega^2}{1-a_1^2}-\frac{2u_\star m_2}{(u_0-u_\star)} 
     +\frac{2(u_\star-1)\mu^3\omega}{(1-a_1^2)(1-a_2^2)(u_0-u_\star)} \\
     +\frac{2(u_\star-1)a_1^2a_2^2\tilde{\Omega}}{\mu(1-a_1^2)
       (1-a_2^2)(u_0-u_\star)}
     -\frac{2(u_\star-1)\mu((a_1^2+a_2^2)\omega-
       a_1(1-a_1^2)m_1-a_2(1-a_2^2)m_2)}{(1-a_1^2)(1-a_2^2)(u_0-u_\star)} \\
     + C_m+\frac{(u_0-u_\star)}{(u_0-1)(u_\star-1)}C_m,
   \end{multline}
   \begin{multline}
     4u_{\star}Q_\star=-\frac{2u_\star(1-\theta_1)}{(u_\star-1)}-
     2m_1+\frac{2u_\star m_2}{(u_0-u_\star)}-
     \frac{2(u_\star-1)\mu^3\omega}{(1-a_1^2)(1-a_2^2)(u_0-u_\star)} \\
     -\frac{2(u_\star-1)a_1^2a_2^2\tilde{\Omega}}{
       \mu(1-a_1^2)(1-a_2^2)(u_0-u_\star)}+
     \frac{2(u_\star-1)\mu((a_1^2+a_2^2)\omega-
       a_1(1-a_1^2)m_1-a_2(1-a_2^2)m_2)}{(1-a_1^2)(1-a_2^2)(u_0-u_\star)}.
  \end{multline}
\end{subequations}
We see again that the position of the apparent singularity at
$u=u_\star$ is related to the parameter $\mu$ through a Möbius
transformation \eqref{eq:ustar} and that the single monodromy parameters
$m_1,m_2,\theta_1$ and $\varsigma = \omega+a_1m_1+a_2m_2$ do not 
depend on $\mu$. The initial proposal of $\mu$-separability
in \cite{Lunin:2017drx} generated some discussion about the
interpretation of the parameter
\cite{Frolov:2018ezx,Frolov:2018eza}. In order to add to that, we need 
to take a detour and write about isomonodromy.
\section{Conditions on the Painlevé VI system}
The most natural setting to describe isomonodromy is the theory of
flat holomorphic connections. The exposition here follows the
monograph by Iwasaki \textit{et al.}, \cite{Iwasaki:1991}, with some
additions suited to our purposes. Consider the matricial system of
differential equations on a single complex variable 
\begin{equation}
  \frac{d\Phi(z)}{dz}
  =\left(\frac{A_0}{z}+\frac{A_1}{z-1}+\frac{A_t}{z-t}\right)\Phi(z),
  \quad\quad A_\infty = -A_0-A_1-A_t=\begin{pmatrix}
    \kappa_1 & 0 \\
    0 & \kappa_2
  \end{pmatrix}.
  \label{eq:matricialsystem}
\end{equation}
Choosing $A_\infty$ diagonal comes at the expense of fixing a basis
for the fundamental solution $\Phi(z)$.  Let us parametrize it as
\begin{equation}
  \Phi(z) = \begin{pmatrix}
    y_1(z) & y_2(w) \\
    w_1(z) & w_2(w)
  \end{pmatrix}.
\end{equation}
It is a straightforward exercise to see that the differential equation
satisfied by the first row of $\Phi(z)$ is
\begin{equation}
  y_i''(z)-\left(\Tr A(z)+\frac{A'_{12}(z)}{A_{12}(z)}\right)y_i'(z)+
  \left(\det
    A(z)-A_{11}'(z)+A_{11}(z)\frac{A_{12}'(z)}{A_{12}(z)}\right)y_i(z)=0
  \label{eq:deformedheun}.
\end{equation}
Furthermore, with $A_\infty$ diagonal, we have
\begin{equation}
  A_{12}(z)=\frac{(A_0)_{12}}{z}+\frac{(A_1)_{12}}{z-1}
  +\frac{(A_t)_{12}}{z-t}=\frac{k(z-\lambda)}{z(z-1)(z-t)},
\end{equation}
so, for the matricial system \eqref{eq:matricialsystem}, the
associated scalar equation \eqref{eq:deformedheun} will be
Fuchsian, with 5 singular points at $z=0,1,t,\lambda,\infty$, exactly
the type encountered in the radial and angular systems, given by
equations \eqref{eq:deformedradial} and \eqref{eq:deformedangular},
respectively.

From this formulation it seems clear that, from the matricial system
\eqref{eq:matricialsystem} perspective, the singularity at $z=\lambda$
is a consequence of our choice of gauge
$A_\infty=\mathrm{diag}(\kappa_1,\kappa_2)$. As a matter of fact, we can see
that there is a residual gauge symmetry that moves $\lambda$, as
discovered by Jimbo, Miwa and Ueno \cite{Jimbo:1981-2}. Let
us introduce the parametrization for the coefficient
matrices $A_i$
\begin{equation}
  A_i=\begin{pmatrix}
    p_i+\hat{\theta}_i & -q_ip_i \\
    \frac{p_i+\hat{\theta}_i}{q_i} & -p_i
  \end{pmatrix},
\end{equation}
which is the most general for the gauge choice where $\Tr
A_i=\hat{\theta}_i$ and $\det A_i=0$, $i=0,1,t$ are fixed. The parameters
$p_i,q_i$ are subject to extra constraints. The diagonal
terms of $A_\infty=-(A_0+A_1+A_t)$ are, 
\begin{equation}
  \kappa_1=\frac{\hat{\theta}_\infty-\hat{\theta}_0-\hat{\theta}_1-\hat{\theta}_t}{2},\quad\quad
  \kappa_2=\frac{-\hat{\theta}_\infty-\hat{\theta}_0-\hat{\theta}_1-\hat{\theta}_t}{2}.
\end{equation}
Let us now define, along with $\lambda$,
\begin{equation}
  \eta = A_{11}(z=\lambda)=\frac{p_0+\hat{\theta}_0}{\lambda}
  +\frac{p_1+\hat{\theta}_1}{\lambda-1}+\frac{p_t+\hat{\theta}_t}{\lambda-t},
\end{equation}
We will now solve for $p_i$ and $q_i$ in terms of $\lambda$ and
$\eta$. The solution will also depend on an extra parameter $k$, which
can be made equal to one by conjugation of all the $A_i$ by a diagonal
matrix, the particular value of $k$ will not enter into
\eqref{eq:deformedheun}. The explicit solutions for $p_i$ and $q_i$
are given as \cite{Jimbo:1981-2}
\begin{equation}
  q_0 = \frac{k\lambda}{tp_0},\quad\quad
  q_1=-\frac{k(\lambda-1)}{(t-1)p_1},\quad\quad
  q_t=\frac{k(\lambda-t)}{t(t-1)p_t},
\end{equation}
with $k$ undefined and
\begin{subequations}
\begin{multline}
  p_0=\frac{\lambda}{t\hat{\theta}_\infty}\left(
    \lambda(\lambda-1)(\lambda-t)\tilde{\eta}^2
    +(\hat{\theta}_1(\lambda-t)+t\hat{\theta}_t(\lambda-1)-2\kappa_2(\lambda-1)
    (\lambda-t))\tilde{\eta}\right.\\
  \left. +\kappa_2^2(\lambda-t-1)-\kappa_2(\hat{\theta}_1+
    t\hat{\theta}_t)\right),
\end{multline}
\begin{multline}
  p_1=-\frac{\lambda-1}{(t-1)\hat{\theta}_\infty}\left(
    \lambda(\lambda-1)(\lambda-t)\tilde{\eta}^2
    +((\hat{\theta}_1+\hat{\theta}_\infty)(\lambda-t)+t\hat{\theta}_t(\lambda-1)
    \right.\\\left.-2\kappa_2(\lambda-1)
    (\lambda-t))\tilde{\eta} +\kappa_2^2(\lambda-t)-\kappa_2(\hat{\theta}_1+
    t\hat{\theta}_t)-\kappa_1\kappa_2\right),
\end{multline}
\begin{multline}
  p_t=\frac{\lambda-t}{t(t-1)\hat{\theta}_\infty}\left(
    \lambda(\lambda-1)(\lambda-t)\tilde{\eta}^2
    +(\hat{\theta}_1(\lambda-t)+t(\hat{\theta}_t+\hat{\theta}_\infty)(\lambda-1)
    \right.\\\left. -2\kappa_2(\lambda-1)
    (\lambda-t))\tilde{\eta}
    +\kappa_2^2(\lambda-1)-\kappa_2(\hat{\theta}_1+
    t\hat{\theta}_t)-t\kappa_1\kappa_2\right),
\end{multline}
\begin{equation}
  \tilde{\eta}=\eta-\frac{\hat{\theta}_0}{\lambda}-\frac{\hat{\theta}_1}{\lambda-1}-
  \frac{\hat{\theta}_t}{\lambda-t}.
\end{equation}
\end{subequations}
In terms of $\lambda$ and $\eta$, the equation \eqref{eq:deformedheun}
is written as
\begin{equation}
\begin{gathered}
  y_{i}''(z)+p(z)y_i'(z)+q(z)y_i(z)=0, \\
  p(z)=\frac{1-\hat{\theta}_0}{z}+\frac{1-\hat{\theta}_1}{z-1}+
  \frac{1-\hat{\theta}_t}{z-t}-\frac{1}{z-\lambda},\\
  q(z) = \frac{\kappa_1(\kappa_2+1)}{z(z-1)}+\frac{t(t-1)K}{z(z-1)(z-t)}
  +\frac{\lambda(\lambda-1)\eta}{z(z-1)(z-\lambda)},
  \label{eq:deformedheun1}
\end{gathered}
\end{equation}
with $\lambda$ and $\eta$ as above and
\begin{equation}
  K=-H-\frac{\lambda(\lambda-1)}{t(t-1)}\eta
  -\frac{\lambda-t}{t(t-1)}\kappa_1
  +\frac{\hat{\theta}_0\hat{\theta}_t}{2t}+\frac{\hat{\theta}_1\hat{\theta}_t}{2(t-1)},
  \label{eq:ktoh}
\end{equation}
where $H$ will be relevant to us in the following
\begin{equation}
  H=\frac{1}{t}\Tr(A_0A_t)+\frac{1}{t-1}\Tr(A_1A_t)
  -\frac{\hat{\theta}_0\hat{\theta}_t}{2t}-\frac{\hat{\theta}_1\hat{\theta}_t}{2(t-1)}.
\end{equation}
We now note that the singularity at $z=\lambda$ in \eqref{eq:deformedheun}
is apparent, and then $K$, $\lambda$ and $\eta$ satisfy an algebraic
constraint. Translating this constraint to $H$,
\begin{equation}
  H=\frac{\lambda(\lambda-1)(\lambda-t)}{t(t-1)}\left(
    \eta^2-\left(\frac{\hat{\theta}_0}{\lambda}+\frac{\hat{\theta}_1}{\lambda-1}
      +\frac{\hat{\theta}_t}{\lambda-t}\right)\eta
    +\frac{\kappa_1\kappa_2}{\lambda(\lambda-1)}\right)\
  +\frac{\hat{\theta}_0\hat{\theta}_t}{2t}+\frac{\hat{\theta}_1\hat{\theta}_t}{2(t-1)}.
  \label{eq:htolambda}
\end{equation}

From the gauge field perspective, we can think of
$A(z)=[\partial_z\Phi(z)]\Phi(z)^{-1}$ as a flat connection, whose
observables are traces of non-contractible Wilson loops. In the
language of complex analysis, holonomy is represented by monodromy
matrices, $M_0$, $M_t$, $M_1$ and $M_\infty$ associated to loops
around each singular point of the systrem \eqref{eq:matricialsystem}.
These matrices are defined up to conjugation and constrained by the
fact that the composition of the monodromies over all singular points
is a contractible curve: 
\begin{equation}
  M_\infty M_1 M_t M_0 = \mathbbold{1}.
\end{equation}
The gauge-invariant observables are the traces of the matrices $\Tr
M_i=2\cos\pi\hat{\theta}_i$\footnote{Here, we have performed a
  normalization of the solution $\Phi(z)$ so that its determinant is
  constant equal to one. One can check that this just subtracts from
  each coefficient matrix $A_i$ its trace.}, and the traces of two of
the composite monodromies:
\begin{equation}
  \Tr M_0M_t=2\cos\pi\sigma_{0t},\qquad
  \Tr M_tM_1=2\cos\pi\sigma_{1t}.
  \label{eq:compositemonodromy}
\end{equation}
The third combination $\Tr M_0M_1$ is related to these two by a
polynomial identity (Fricke-Jimbo relation), involving the
$\hat{\theta}_i$, as can be seen in \cite{Jimbo:1982aa}.

The gauge-invariant quantities
$\hat{\theta}_0,\hat{\theta}_t,\hat{\theta}_1,
\hat{\theta}_\infty,\sigma_{0t},\sigma_{1t}$ are called
\textit{monodromy data}, associated to the system
\eqref{eq:matricialsystem} -- or,  alternatively, to the equation
\eqref{eq:deformedheun}. Of those, the single monodromy parameters
$\hat{\theta}_i$ can be read directly from the differential equation,
whereas $\sigma_{0t}$ and $\sigma_{1t}$ are not readily available. On
the other hand, they comprise the information needed from the
equation to solve the scattering problem \cite{Novaes:2014lha}, or to
find quasinormal modes \cite{Barragan-Amado:2018pxh}. Therefore,
finding them is a problem of interest.

Solving for $\sigma_{0t},\sigma_{1t}$ makes use of a residual gauge
symmetry of \eqref{eq:matricialsystem}, which changes the position of
the singularity at $z=t$. The zero curvature condition
$\partial_z\partial_t\Phi(z,t)=\partial_t\partial_z\Phi(z,t)$ forces
the coefficient matrices $A_i$ to satisfy the \textit{Schlesinger
  equations} 
\begin{gather}
  \frac{\partial A_0}{\partial t} = -\frac{1}{t}[A_0,A_t],\quad\quad
  \frac{\partial A_1}{\partial t}=-\frac{1}{t-1}[A_1,A_t],\nonumber\\
  \frac{\partial A_t}{\partial
    t}=\frac{1}{t}[A_0,A_t]+\frac{1}{t-1}[A_1,A_t],
\end{gather}
which, when written in terms of $\lambda$ and $\eta$, result in the
Painlevé VI transcendent. 

In a seminal paper \cite{Jimbo:1982aa}, Jimbo derived asymptotic
expansions for the Painlevé VI transcendent in terms of monodromy
data, written in a slightly different guise: 
\begin{equation}
  \frac{\partial}{\partial
    t}\log\tau(\hat{\vec{\theta}},\vec{\sigma};t)=H
  -\frac{1}{2t}\hat{\theta}_0\hat{\theta}_t-\frac{1}{2(t-1)}\hat{\theta}_1\hat{\theta}_t,
\end{equation}
where $\tau$ is a function of the monodromy data and of $t$, called the
isomonodromic time. In another big development, \cite{Gamayun:2012ma}
gave the full expansion for $\tau$, given generic monodromy
arguments, in terms of Nekrasov functions, with the structure
\begin{equation}
  \tau(\hat{\vec{\theta}},\vec{\sigma};t) = \sum_{n\in\mathbb{Z}}
  C(\hat{\vec{\theta}},\sigma_{0t}+2n) [s(\hat{\vec{\theta}},\vec{\sigma})]^n
  t^{\frac{1}{4}\sigma_{0t}^2+n(\sigma_{0t}+n)}
  \mathcal{B}(\hat{\vec{\theta}},\sigma_{0t}+2n;t),
  \label{eq:nekrasov}
\end{equation}
where the Nekrasov functions $\mathcal{B}$ are analytic in $t$. We refer to
\cite{Amado:2017kao} for details. These functions were introduced as
the instanton partition function of four-dimensional ${\cal N}=2$
$\mathrm{SU}(N)$ Yang-Mills \cite{Nekrasov:2002qd} coupled to matter
multiplets, and were related to two dimensional conformal blocks by
the Alday-Gaiotto-Tachikawa conjecture \cite{Alday:2009aq}, later
proved by Alba, Fateev, Litvinov and Tarnopolsky
\cite{Alba:2010qc}. The relation then comes full circle to 
help solve classical field propagation in five dimensional space-times.

With the full expansion given in \cite{Gamayun:2012ma,Gamayun:2013auu},
connection formulas for the expansions at different singular points
were given \cite{Iorgov:2013uoa}, and a Fredholm determinant
formulation was constructed \cite{Gavrylenko:2016zlf}, which is
well-suited to numerical calculations \cite{Anselmo:2018}.

The monodromy problem, which is the original formulation of the
classical Riemann-Hilbert problem, consists in finding the full set of
monodromy data from the parameters in the equation
\eqref{eq:deformedheun1}. For our purposes, the latter will consist of
the single monodromy data $\{\hat{\theta}_i\}$ and the parameters
$\lambda,\eta$ -- remember that $K$ is related to them by
\eqref{eq:ktoh} and \eqref{eq:htolambda}. These conditions are best
written in terms of the $\zeta$ function defined in
\cite{Okamoto:1986aa} (called $\sigma$(t) there), 
\begin{equation}
  \zeta(t) = t(t-1) \frac{\partial}{\partial
    t}\log\tau(\hat{\vec{\theta}},\vec{\sigma};t)
  =(t-1)\Tr A_0A_t+t\Tr A_1A_t
  -\frac{t-1}{2}\hat{\theta}_0\hat{\theta}_t-\frac{t}{2}
  \hat{\theta}_1\hat{\theta}_t.
  \label{eq:zetafcn}
\end{equation}
In terms of $\zeta(t)$, the Schlesinger equations read
\begin{equation}
  \frac{d\zeta}{dt}(t)=-\Tr(A_t(A_t+A_\infty))-\frac{1}{2}
  (\hat{\theta}_0+\hat{\theta}_1)\hat{\theta}_t,\qquad
  \frac{d^2\zeta}{dt^2}(t)=-\frac{\Tr(A_\infty[A_0,A_t])}{t(t-1)}.
  \label{eq:zetaderivatives}
\end{equation}

The strategy to recover the monodromy data from
\eqref{eq:deformedheun1} is now clear: given the differential
equation, which is parametrized by particular values for
$\lambda_0,\eta_0$, and the monodromy time $t_0$, one can find the
coefficient matrices $A_i$ -- up to overall conjugation -- by solving
for $p_i$'s and $q_i$'s using the formula above. Given \eqref{eq:ktoh}
and \eqref{eq:htolambda}, as well as the formulas for the entries of
$A_i$ above, we can then compute the derivatives of $\zeta$ 
at $t=t_0$
\begin{subequations}
\begin{multline}
  \zeta(t_0)=\lambda_0(\lambda_0-1)(\lambda_0-t_0)
  \left(\eta_0^2-\left(\frac{\hat{\theta}_0}{\lambda_0}
      +\frac{\hat{\theta}_1}{\lambda_0-1}+\frac{\hat{\theta}_t}{\lambda_0-t_0}
    \right)\eta_0+\frac{\kappa_1\kappa_2}{\lambda_0(\lambda_0-1)}
  \right)\\
  +\frac{t-1}{2}\hat{\theta}_0\hat{\theta}_t+\frac{t}{2}\hat{\theta}_1\hat{\theta}_t,
\end{multline}
\begin{multline}
  \frac{d\zeta}{dt}(t_0)=-\frac{\lambda_0(\lambda_0-1)(\lambda_0-t_0)^2
  }{t_0(t_0-1)}\left(\eta_0^2-\left(\frac{\hat{\theta}_0}{\lambda_0}+
      \frac{\hat{\theta}_1}{\lambda_0-1}+\frac{\hat{\theta}_t
        -\hat{\theta}_\infty}{\lambda_0-t_0} 
   \right)\eta_0
   +\frac{\kappa_1^2}{(\lambda_0-t_0)^2}
 \right)\\-\frac{\lambda_0-1}{t_0-1}\kappa_1\hat{\theta}_0
 -\frac{\lambda_0}{t_0}\kappa_1\hat{\theta}_1
 -\kappa_1\kappa_2+\frac{1}{2}(\hat{\theta}_0+\hat{\theta}_1)\hat{\theta}_t,
\end{multline}
\label{eq:zetainitial}
\end{subequations}
and the second derivative can be written in terms of $\zeta(t)$ and
$\zeta'(t)$,  resulting in the ``$\sigma$-form'' of the Painlevé VI equation
\begin{multline}
  \zeta'(t(t-1)\zeta'')^2+[2\zeta'(t\zeta'-\zeta)-(\zeta')^2
  -\tfrac{1}{16}(\hat{\theta}_t^2-\hat{\theta}_\infty^2)
  (\hat{\theta}_0^2-\hat{\theta}_1^2)]^2\\
  =(\zeta'+\tfrac{1}{4}(\hat{\theta}_t+\hat{\theta}_\infty)^2)
  (\zeta'+\tfrac{1}{4}(\hat{\theta}_t-\hat{\theta}_\infty)^2) 
  (\zeta'+\tfrac{1}{4}(\hat{\theta}_0+\hat{\theta}_1)^2)
  (\zeta'+\tfrac{1}{4}(\hat{\theta}_0-\hat{\theta}_1)^2).
  \label{eq:zetapainleve}
\end{multline}

In principle, the initial conditions for $\zeta(t)$ at $t=t_0$ above
determine $\zeta(t)$ uniquely through the differential
equation. The function can then be inverted to recover $\sigma_{0t}$
and $\sigma_{1t}$.

We note that the change of the parameters $\lambda$ and $\eta$ with
respect to $t$, along the isomonodromy solution is a gauge
transformation in the sense that the monodromy data is kept invariant,
therefore
\begin{equation}
  \delta \lambda = \frac{\partial K}{\partial \eta}\delta t,\qquad
  \delta \eta = -\frac{\partial K}{\partial \lambda}\delta t,
  \label{eq:hamiltonian}
\end{equation}
is a residual gauge transformation.

\section{Formal solution to the radial and angular systems}
\label{sec:formal}

\subsection{Writing the boundary conditions in terms of monodromy
  data}

In terms of the Painlev\'e VI $\tau$-function, or rather the $\zeta$
function defined in \eqref{eq:zetafcn}, the parameters of
ODE comprise an initial value problem for the $\zeta$ function. The
case of angular and radial equations above is a little more involved
since the parameters are coupled. Let us start with the following
identification 
\begin{center}
  \label{tab:parameters}
  \begin{tabular}{|c|c|c|c|c|c|c|c|c|}
     \hline & $t_0$ & $\hat{\theta}_{0}$ & $\hat{\theta}_{t}$ & $\hat{\theta}_{1}$ &
      $\hat{\theta}_{\infty}$ & $K$ & $\lambda_0$ & $\eta_0$  \\  \hline
      $\zeta_{\text{Rad}}(t)$& $z_0$ & $\theta_{-}$ & $\theta_{+}$ &
      $\theta_1$ & $\theta_{0} $ & $K_0(\mu,C_m)$ & $z_{\star}(\mu)$ &
      $K_\star(\mu)$ \\  \hline
      $\zeta_{\text{Ang}}(t)$& $u_0$ & $-m_1$ & $-m_2$ & $\theta_1$ &
      $\varsigma$ & $Q_0(\mu,C_m)$ & $u_{\star}(\mu)$ & $Q_\star(\mu)$ \\ 
    \hline
  \end{tabular}
\end{center}
where we highlighted the dependence of the parameters of the radial
and angular systems on $\mu$ and the separation
constant $C_m$. It is a straightforward exercise to show that the
corresponding quantities $K$, $\lambda_0$, $\eta_0$ and $t_0$ for the
radial and angular systems are not independent, satisfying
\eqref{eq:ktoh} and \eqref{eq:htolambda}. This fact shows that the
singularity at $\lambda_0$ for both radial and angular systems is
apparent, as anticipated in Sec. \ref{sec:maxwelltoheun}. 

For the angular system, we want to solve the eigenvalue problem. So,
in principle, \eqref{eq:zetainitial} gives a condition on the generic
solution of the Painlevé equation \eqref{eq:zetapainleve}, given by
\eqref{eq:nekrasov}, in which we read the two monodromy parameters
\begin{equation}
  \sigma_{0,u_0;\mathrm{Ang}}(\omega,\mu,C_m),\qquad\text{and}\qquad
  \sigma_{u_0,1;\mathrm{Ang}}(\omega,\mu,C_m),
\end{equation}
where we omitted the dependence on $m_1,m_2,a_1,a_2$. 

As discussed in \cite{Amado:2017kao,Barragan-Amado:2018pxh}, the
condition that the solutions of the angular differential equation
\eqref{eq:angularheun} are well-behaved both at the North and South
poles of the sphere $x,\phi,\psi$ can be written in terms of the
monodromy parameters. Let us now review this construction.

Let $y_{1,2;i}(z)$ be (normalized) solutions of the deformed Heun equation
\eqref{eq:deformedheun} associated to a fundamental matrix $\Phi_i(z)$
whose monodromy matrix is diagonal at a chosen regular singular point $z=z_i$:
\begin{equation}
  \Phi_i((z-z_i)e^{2\pi i}+z_i)=\Phi_i(z)e^{\pi
    i\hat{\theta}_i\sigma_3}.
\end{equation}
Note that this implies that the solutions $y_{1,2;i}(z)$ have
different behavior asymptoting $z_i$, with $y_{1;i}(z)/y_{2;i}(z)\propto
(z-z_i)^{\theta_i}$ as $z\rightarrow z_i$.
The analogous solution at a different singular point $z=z_j$,
$\Phi_j(z)$, is connected to $\Phi_i(z)$ by a constant matrix $E_{ij}$
\begin{equation}
  \Phi_j(z)=\Phi_i(z)E_{ij},
\end{equation}
called the \textit{connection matrix} between $z_i$ and $z_j$. It is
straightforward to see that, if a given solution $y(z)$ of
\eqref{eq:deformedheun} has a definite behavior, in
the sense that it asymptotes one of the solutions at $z=z_i$, say
$y_{1,i}(z)$, and one of the solutions at $z=z_j$, say $y_{1,j}(z)$,
then the connection matrix $E_{ij}$ must be triangular. This in turn
implies that the monodromy matrix of $\Phi_i(z)$ around $z=z_j$,
generically of the form $M_j=E_{ij}e^{\pi
  i\hat{\theta}_j\sigma_3}E_{ij}^{-1}$, is also triangular, and then
the composite monodromy parameter $\sigma_{ij}$ will satisfy
\begin{equation}
  2\cos\pi\sigma_{ij}=\Tr
  M_iM_j=2\cos\pi(\hat{\theta}_i+\hat{\theta}_j) \longrightarrow
  \sigma_{ij}=\hat{\theta}_i+\hat{\theta}_j+2m,\quad m\in\mathbb{Z}.
  \label{eq:quantizationsigma}
\end{equation}
It is also straightforward to show that the converse is also true: if
the composite monodromy $\sigma_{ij}$ satisfies
\eqref{eq:quantizationsigma}, then the connection matrix is
triangular.

Coming back to the angular system, the condition that the solutions
are well-behaved at the North pole ($x=a_2^2$, or $u=0$) and at the
South pole ($x=a_1^2$, or $u=u_0$) means
\begin{equation}
  \sigma_{0,u_0;\mathrm{Ang}}(\omega,\mu,C_m)=-m_1-m_2-2m,
  \quad m\in\mathbb{Z}
  \label{eq:separationconstant}
\end{equation}
which defines the separation constant as an integer
family of functions of $\mu$ and $\omega$: $C_m(\mu,\omega)$. We will
overlook issues of existence and uniqueness for the purposes of this
exposition.

Now, for the radial system, again the solution of the isomonodromic
equation \eqref{eq:zetapainleve} with the initial conditions
\eqref{eq:zetainitial} will define the corresponding two composite monodromy
parameters, associated to paths encircling the singularities at
$z=0,z_0$ and $z=z_0,1$, respectively 
\begin{equation}
  \sigma_{0,z_0;\mathrm{Rad}}(\omega,\mu,C_m),\qquad
  \sigma_{z_0,1;\mathrm{Rad}}(\omega,\mu,C_m).
  \label{eq:radialquantization}
\end{equation}
where we omit the dependence on the other physical parameters
$M,a_1,a_2,m_1,m_2$. It is customary to substitute in this condition
the expression for the separation constant $C_m$ obtained from the
angular equation. We will postpone this step for now.

As an aside, note that the interest in the radial and angular systems,
apart from finding the actual form of the radial wavefunctions -- whose
local expansions can be obtained from Frobenius method -- usually
consists of the scattering problem and the quasinormal modes problem.
They can both be cast in terms of the monodromy parameters
problem, with now the relevant composite monodromy parameter involving
the outer horizon $r=r_+$, or $z=z_0$ and the conformal boundary at
$r=\infty$, or $z=1$. The transmission coefficient, for instance, is
\cite{daCunha:2015fna} 
\begin{equation}
  |\mathcal{T}|^2=\left|\frac{\sin\pi\theta_+\sin\pi\theta_0}{
      \sin\frac{\pi}{2}(\sigma_{z_0,1;\mathrm{Rad}}-\theta_++\theta_1)
      \sin\frac{\pi}{2}(\sigma_{z_0,1;\mathrm{Rad}}+\theta_+-\theta_1)}
  \right|,
  \label{eq:transmcoeff}
\end{equation}
which poses an interpretation problem for $\mu$: since in principle the
electric and magnetic modes \eqref{eq:ansatz} exhaust the $3$
polarizations of the photons in 
five dimensions -- with $1$ of them electric and $2$ magnetic as
argued in \cite{Lunin:2017drx} -- the fact that the scattering
coefficient depends on the extra parameter $\mu$ seems spurious.

This redundancy also arises in the calculation of quasinormal modes from the radial system,
whose method of solution mirrors that of the angular eigenvalue. The requirement that the radial
wavefunction is ``purely outgoing'' at $r=r_+$ and ``purely ingoing''
at $r=\infty$ can be cast, by the same arguments put forward above, in
terms of the quantization condition for the composite monodromy
parameter 
\begin{equation}
  \sigma_{z_0,1;\mathrm{Rad}}(\omega,\mu)=\theta_+(\omega)-\sqrt{1-
    C_m(\omega,\mu)}+2n,
  \qquad n\in\mathbb{Z}.
  \label{eq:eigenfreqs}
\end{equation}
This condition defines implicitly the modes $\omega_{n,m}(\mu)$ as a
function of the radial and azimuthal quantum numbers $n$ and
$m$. Again, it seems rather unphysical that there will be a
1-parameter family of vector quasinormal modes in the space-time.

The redundancy is solved by the intertwining of both radial
and angular systems. Note
that, in both angular and radial equations, \eqref{eq:deformedangular}
and \eqref{eq:deformedradial}, the function that parametrizes the
position of the apparent singularity, represented by $\lambda$ in the
generic deformed Heun equation \eqref{eq:deformedheun1} is
essentially, up to a global conformal transformation, the parameter
$\mu$: 
\begin{equation}
  \lambda_{\mathrm{Ang}}=u_\star=\frac{\mu^2-a_1^2}{\mu^2-1},
  \qquad
  \lambda_{\mathrm{Rad}}=z_\star=\frac{r_-^2+\mu^2}{r_0^2+\mu^2}.
\end{equation}
Using the properties of the Painlev\'e VI equation, the function
$\lambda$ can be computed using the $\zeta$ function defined in
\eqref{eq:zetafcn} by resorting to the demonstration in
\cite{Okamoto:1986aa},  
\begin{multline}
  \frac{1}{\lambda-t}=-\frac{1}{2}\left(\frac{1}{t}+\frac{1}{t-1}\right)
  \\
  +\frac{\hat{\theta}_\infty
    t(t-1)\zeta''+(\zeta'+\frac{1}{4}(\hat{\theta}_t^2-\hat{\theta}_\infty^2))
    ((2t-1)\zeta'-2\zeta+\frac{1}{4}(\hat{\theta}_0^2-\hat{\theta}_1^2))+
    \frac{1}{4}\hat{\theta}_\infty^2(\hat{\theta}_0^2-\hat{\theta}_1^2)}{
    2t(t-1)(\zeta'+\frac{1}{4}(\hat{\theta}_t-\hat{\theta}_\infty)^2)
    (\zeta'+\frac{1}{4}(\hat{\theta}_t+\hat{\theta}_\infty)^2)},
  \label{eq:okamotorel}
\end{multline}
using the Hamiltonian properties of the isomonodromic system. Equation
\eqref{eq:okamotorel}, as well as analogues for the other Painlevé 
transcendents, can be found in \cite{Gamayun:2012ma}.

With the help of \eqref{eq:okamotorel}, we can lift the
ambiguity, using the procedure we can now describe. The angular and
radial systems define four monodromy parameters as stated
above. Of these, the quantization condition for the angular solutions
and the quasinormal modes will set two,
\begin{equation}
  \sigma_{0,u_0;\mathrm{Ang}}(\omega,\mu,C_m),\qquad
  \sigma_{z_0,1;\mathrm{Rad}}(\omega,\mu,C_m)
\end{equation}
which can be used to implicitly define $\omega(\mu)$ and $C_m(\mu)$ as
functions of the redundancy parameter $\mu$. Now, with this
substitution, one can write the two remaining monodromy parameters as
functions of $\mu$
\begin{equation}
  \sigma_{u_0,1;\mathrm{Ang}}(\omega(\mu),\mu,C_m(\mu)),\qquad
  \sigma_{0,z_0;\mathrm{Rad}}(\omega(\mu),\mu,C_m(\mu)).
\end{equation}
These four one-parameter families of monodromy parameters can now be
fed into \eqref{eq:nekrasov} to define two one-parameter families of
$\zeta$ functions, one for the angular system and one for the radial
system. Calling $\lambda_{\mathrm{Ang}}(\mu)$ and
$\lambda_{\mathrm{Rad}}(\mu)$ the respective functions defined by the
right-hand side of \eqref{eq:okamotorel}, we find an extra condition
by requiring that the parameter $\mu$ defined by both systems is
equal:
\begin{equation}
  \mu^2=
  \frac{\lambda_{\mathrm{Ang}}(\mu)-a_1^2}{
    \lambda_{\mathrm{Ang}}(\mu)-1}
  =\frac{r_-^2-\lambda_{\mathrm{Rad}}(\mu)r_0^2}{
    \lambda_{\mathrm{Rad}}(\mu)-1},
  \label{eq:okamotocond}
\end{equation}
which can be seen as a consistency condition for both
isomonodromic systems defined by the angular and radial equations. One
can rephrase this property in different ways, and we will find below
that using the expression for the second derivative of $\zeta$ given
by \eqref{eq:zetaderivatives} as a fifth condition, along with the
values of the monodromy parameters above is a more computationally
efficient approach. This condition, along with the corresponding
quantization conditions for the angular \eqref{eq:separationconstant}
and radial system \eqref{eq:eigenfreqs} are sufficient to determine
all the separation constants, as well as the frequencies for the
quasi-normal modes. This procedure is in stark contrast to the role of
$\mu$ in the four-dimensional case studied previously
\cite{Dolan:2018dqv}, where it can be eliminated by a simple change of
variables. 

We should point out, however, that the implicit definitions for the
quantities $\omega, C_m,\mu$ presented above may not be single
valued, which will then allow for orbits of $\mu$ with disconnected
components. In particular, one can indeed have more than one solution
to \eqref{eq:okamotocond}. In the next section, we will take the
solution closest to $u_0$, due to the nature of the Nekrasov
expansion. It is an open question whether a different choice will lead
to different physics. We will leave the study of these subtleties for
future work. 

\subsection{Quasinormal modes from the radial system}

The conditions put forth in the last Section provide an exact,
procedural solution to the quasinormal modes for the vector
perturbations. However, the analytical treatment of these conditions
is of very limited scope at this moment, given the five transcendental
equations one has to deal with. One may, however, resort to numerical
implementations of the $\tau$ function.

In this Section we are going to describe  the numerical treatment of
eigenfrequencies for ultraspinning $a_1\rightarrow 1$ black
holes. This amounts to solving numerically in this limit the equations
listed previously for  $\{\omega, \sigma_{0,z_0;Rad},
\sigma_{z_0,1;Rad}, \mu, C_{m}, \sigma_{u_0,1;Ang}, 
\sigma_{0,u_0;Ang}\}$.  For $a_1\rightarrow 1$, the angular system is
better served by the expansion of the $\tau$ function around $u_0=1$,
which was given in \cite{Gamayun:2012ma}. Recently, in
\cite{Cayuso:2019ieu}  a similar analysis was performed for four
dimensional Kerr-Newman and Kerr-Sen black holes. Nevertheless, our
problem presents an extra difficulty related to the presence of $\mu$
in the radial and angular equations, which is not the case in four
dimensions where the separation parameter $\mu$ can be absorbed by a
redefinition of the separation constant $C_m$, as can be seen in
\cite{Dolan:2018dqv,Frolov:2018ezx}. See also \cite{Cayuso:2019vyh}
for a similar discussion.  

\begin{figure}[htb]
  \begin{center}
  \mbox{\includegraphics[width=0.9\textwidth]{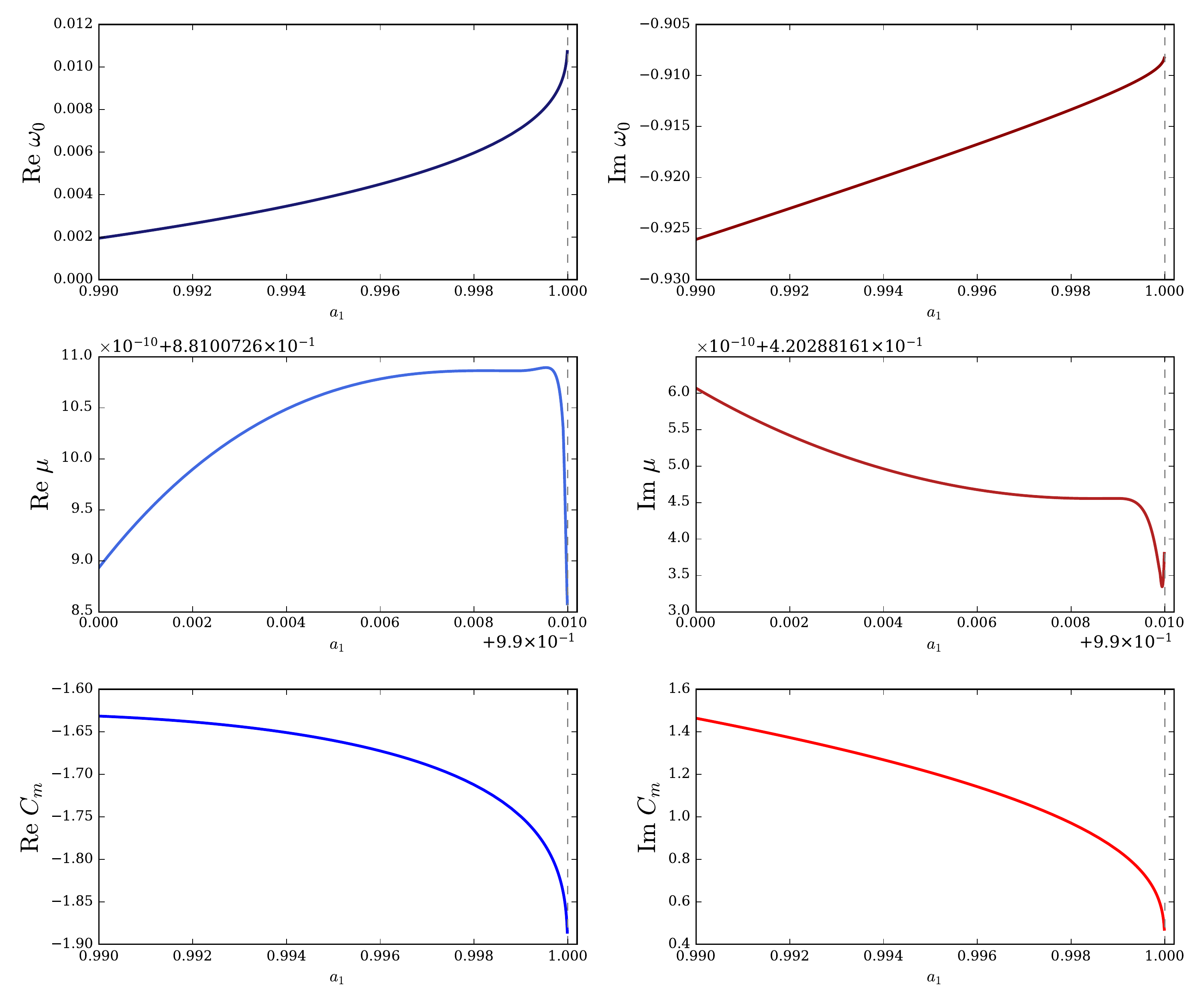}}
  \caption{Quasinormal frequencies $\omega_{0}$, separation constant
    $C_{m}$ and the parameter $\mu$ for the electric polarization in
    the ultraspinning limit. Note that the change of the separation
    parameter $\mu$ is very small, in the range of variation of $a_1$
    considered.} 
  \label{fig:omega_and_Cm}
  \end{center}
\end{figure}

In Fig. \ref{fig:omega_and_Cm}, we display the results, with the
rotation parameter $a_2=0.001$ and the size of the outer horizon 
$r_{+}=0.05$. 
The ultraspinning black hole regime considered is $0.99 \leq
a_{1} \leq 0.99999$, and the values of the quantum numbers are set at $\ell=2,
m_{1}=m_{2}=0$. The frequencies found are stable and increase
monotonically with $a_1$. The value of $\mu$ has a more complicated
behavior with $a_1$, but it should be kept in mind that the change
comes in the tenth decimal place and may be affected by numerical
errors. Even with this caveat, the approximation of the angular $\tau$
function improves as $a_1\rightarrow 1$ and the expansion of the radial
$\tau$ function should be valid as long as $|\theta_+z_0|\ll 1$, with
typical values in the range $|\theta_+z_0|\sim 10^{-2}$. In our
analysis, we have used the Nekrasov expansion \eqref{eq:nekrasov}
truncating the number of channels $n\le N_{c}=7$ and the number of
levels in the conformal block ${\cal B}$ to $m\le N_{b}=7$. For
$z_0\approx 10^{-2}$, we estimate at least 10 digits accuracy.

Note that the quasinormal frequency $\omega_0$ increases until
$a_1=0.999$, where it shows an asymptote, and possibly a non
polynomial behavior in $1-a_1$. A better understanding of the physics
behind this analysis is indeed well-deserved, particularly the 
prospect of superradiant modes \cite{Barragan-Amado:2018pxh}. We will,
however, leave these aspects to future work.

\begin{figure}[htb]
  \begin{center}
  \mbox{\includegraphics[width=0.58\textwidth]{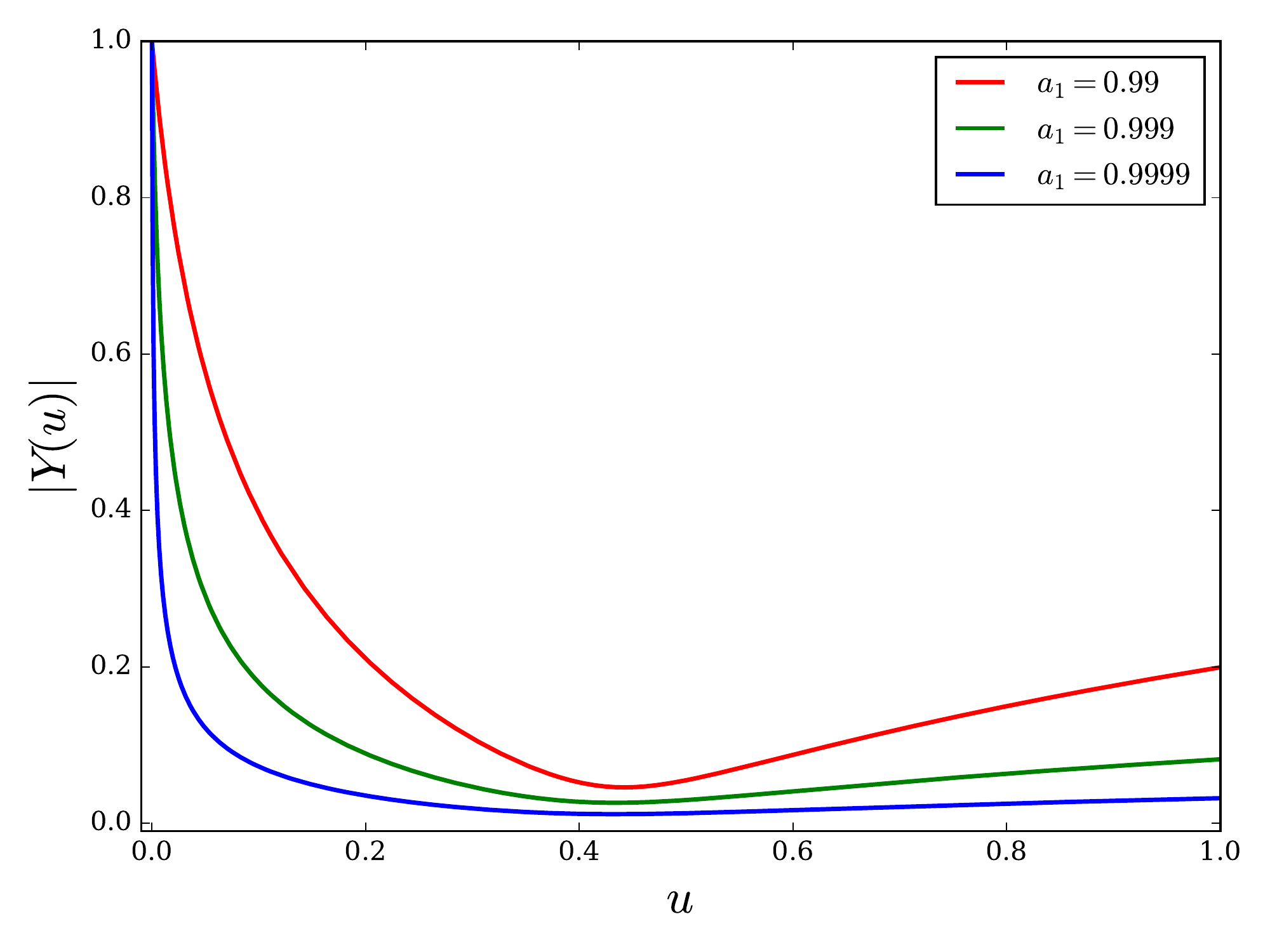}}
  \caption{Numerical eigenfunctions of the angular equation
  \eqref{eq:deformedangular} for different values of $a_1$. These
    were obtained by matching the Frobenius expansion at two of the singular
    points, $u=0$ and $u=u_0$, with $16$ terms. The values of $\mu$ were
    chosen by requiring consistency with the radial system.} 
  \label{fig:angular}
  \end{center}
\end{figure}

As a preliminary test of the results above, we have
constructed the angular eigenfunctions $Y(u)$ using the values for
$\{\omega,C_m,\mu\}$ obtained using isomonodromy for a few values of
$a_1$ and plotted in Fig. \ref{fig:angular}. 
The construction is
the standard Frobenius method, where expansions for $Y(u)$ at both
points $u=0$ and $u=u_0\approx 1$ and matching at middle
point are performed for the value of the
function as well as 15 derivatives. The asymptotic behavior in
Fig. \ref{fig:angular}  is as expected, and we could verify the values
of the parameters obtained to at least 10 digits. Unfortunately, the
construction for the radial eigenfunctions is much more
computationally demanding and a detailed analysis is also postponed to
future work.

\section{Discussion}
In this work we studied the role of the separation parameter $\mu$,
introduced in \cite{Lunin:2017drx} to allow for the separation of
Maxwell equations in a five-dimensional Kerr-(anti) de Sitter
background. We saw that $\mu$ is related to an apparent singularity of
the resulting angular and radial differential equations. Specifically,
the position of the apparent singularity is related to $\mu$ by a
simple Möbius transformation. After translating the boundary
conditions for the radial equation \eqref{eq:eigenfreqs} and angular
equation \eqref{eq:separationconstant} in terms of monodromy data, we
could outline a method to find the separation constant and the
quasinormal modes frequencies. The separation parameter $\mu$ is fixed
by a consistency condition between the $\tau$ functions for the radial
and angular systems \eqref{eq:okamotocond}. We have then checked the 
procedure numerically by considering small $r_+\ll 1$ and
ultraspinning black holes $a_1=0.001,a_2\lesssim 1$.

We should point out now that, unlike the scalar case studied in
\cite{Barragan-Amado:2018pxh}, the use of the $\tau$ function and the
isomonodromy method for the vector case is not just a numerically more
efficient way for computing the quasinormal modes. The monodromy
language allows us to define the quantities involved in a way
independent of $\mu$ and hence to decouple the conditions necessary to
solve the problem. It is interesting to notice that the introduction
of the apparent singularity mirrors the remark by Poincaré that
apparent singularities are necessary in ordinary differential
equations if we want to solve the problem of finding the parameters of
the ordinary differential equation whose solutions are associated to
generic monodromy parameters \cite{Iwasaki:1991}. We hope that the
results in this paper can help to elucidate the geometrical
structure behind the introduction of $\mu$. We also expect that the
method presented here will help with the solution for the Proca and
$p$-form fields in the same background, which were shown to lead to
separable equations in \cite{Lunin:2019pwz}.

One can deduce from the analysis that the separation parameter
$\mu$ in higher dimensions plays a more prominent role than in the four
dimensional case, where it can be eliminated by a suitable change of
variables. The case studied here, that of ``electric'' polarizations,
as defined in \cite{Lunin:2017drx} is related to the ``magnetic''
polarizations by an inversion $\mu\rightarrow 1/\mu$. Although this
inversion could be interpreted as a gauge transformation of the
electric mode, we note that it nevertheless modifies the asymptotic
behavior of the field, so a more careful analysis is in order.
At any rate, the results above found that $\mu$ should assume a
discrete set of values, at least to allow solutions for the
quasi-normal modes. In turn, the latter fact opens up the possibility
of studying the different polarizations by exploring the symmetries of
the Painlevé system. Another outstanding problem is the relation among
the different definitions of polarizations in the literature
\cite{Krtous:2018bvk}. We leave the exploration of these issues for
future work.

The fact that the separation parameter $\mu$ parametrizes
the position of the extra apparent singularity for both the angular and
the radial equations seems an indication that perhaps the matrix
system \eqref{eq:matricialsystem} plays a more prominent role than
previously thought. That the particular choice of parameters,
including $\mu$, is able to decouple the equations mirrors the
treatment of conformal blocks of conformal field theories in higher
($D>2$) dimensions \cite{Poland:2018epd} where there exists a
particular set of coordinates which factorizes the partial wave
expansion as a product of two hypergeometric functions (at least in
$D=4$, see \cite{Dolan:2000ut}).

Given the holographic aspect of perturbations of AdS$_5$ space,
perhaps the last point is more than an analogy. One may hope that the
results here can be of use for the study of conformal bootstrap and
conformal perturbation theory in four dimensions. On a more immediate
direction, a systematic study of the quasinormal modes for generic
black hole parameters as well as the near extremal case, with the
prospect of instabilities, is also necessary.  

\section*{Acknowledgements}
The authors are greatly thankful to Oleg Lunin for triggering our
interest in the separability of Maxwell's equations, and for his
comments on an earlier version of this manuscript. BCdC is also
thankful of the hospitality of the ICTP-Trieste where part of this
work was completed. This work has been partially funded by 
CAPES, CNPq (N$^{\underline{\circ}}$ 302865/2017-9) and the University
of Groningen in the Netherlands. 


\providecommand{\href}[2]{#2}\begingroup\raggedright\endgroup

\end{document}